\documentclass[12pt,a4paper,dvips]{article}
\usepackage{epsfig,wrapfig,times,mathptm} 
\renewcommand{\section}[1]{\vspace{6pt} \noindent\mbox{#1} \newline \noindent}
\renewcommand{\subsection}[1]{\vspace{6pt} \noindent\mbox{\underline{#1}} 
\newline \noindent}
\renewcommand{\subsubsection}[1]{\vspace{6pt} \noindent\mbox{\underline{#1}}
\noindent}

\newfont{\sansb}{cmssbx10}
\newfont{\sans}{cmss10}

\font\it=cmti10 scaled\magstep 1
\begin{document}
{\small 4.1.26 \vspace{-24pt}\\}     
{\center \LARGE STATUS OF RADIO ICE CERENKOV EXPERIMENT (RICE)
\vspace{6pt}\\}
C. Allen$^3$, A. Bean$^2$, D. Besson$^2$, G. Frichter$^1$,
S. Kotov$^2$, I. Kravchenko$^2$,
D. McKay$^2$, T. C. Miller$^1$, L. Piccirillo$^1$, J. Ralston$^2$,
D. Seckel$^1$, S. Seunarine$^2$,
G.M. Spiczak$^1$  \vspace{6pt}\\
{\it $^1$Bartol Research Institute, University of Delaware, Newark, DE  
19716  USA\\
$^2$Department of Physics, University of Kansas, Lawrence, KS  66045  USA\\
$^3$Dept. of Elec. Eng. and Comp. Sci., 
University of Kansas, Lawrence, KS  66045  USA\\
\vspace{-12pt}\\}
{\center ABSTRACT\\}
The Radio Ice Cerenkov Experiment (RICE) is designed to 
detect ultrahigh energy ($\geq$100~TeV) neutrinos from astrophysical
sources.  
RICE will consist of an array of compact
radio (100 to 1000~MHz) receivers buried in ice at the South Pole.
The objective is an array of greater than one cubic
kilometer effective volume, complementary to
TeV optical neutrino telescopes.
The effective volume using the radio technique increases faster
with energy than the optical technique, making the method more
efficient at ultrahigh energies.
During the 1995-96 and 1996-97 austral summers, several receivers
and transmitters were deployed in bore holes drilled for the
AMANDA project, at depths of 141 to 260~m.  This was the
first {\it in situ\/} test of radio receivers in deep ice for neutrino
astronomy.

\setlength{\parindent}{1cm}
\begin{wrapfigure}[20]{r}[0pt]{7.5cm}
\epsfig{file=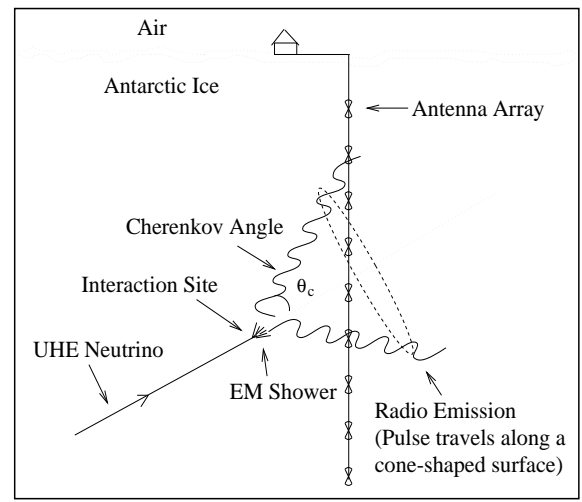,width=7.5cm,height=6.5cm,angle=0}
\caption{\it The RICE concept.  An UHE electron neutrino
initiates an electromagnetic shower in Antarctic ice.  The resulting
radio pulse is detected by a buried array of receivers.}
\end{wrapfigure} 
\section{INTRODUCTION}
The detection of ultrahigh energy neutrinos
represents a unique opportunity in astrophysics.
Photons of such energy are attenuated on the cosmic
microwave background, while protons
are deflected in intergalactic magnetic fields and do not
point back to their source.
The potential to observe objects not seen by
any other method
has stimulated much recent theoretical and experimental work.
Several production sites
of high energy neutrinos have been theorized, including
massive black holes at the center of the Milky Way or
Active Galactic Nuclei (AGN), superconducting cosmic 
strings, the sources of the highest energy cosmic rays,
gamma ray bursts, young supernova remnants, and X-ray binary
systems.
Fluxes from these potential sources are small
enough that detectors with active volumes on
the order of 1 km$^{3}$ are needed in
order to observe them (Gaisser et al., 1994).
Several high energy neutrino projects, 
including AMANDA (see HE 4.1.1),
NESTOR (Resvanis, 1993), and Baikal (Bezrukov, 1995),
are now underway.
These projects all use photomultiplier tubes in deep water
or ice to observe
visible and UV Cerenkov light emitted by muons created 
in charged current interactions by $\nu_{\mu}$,
and all are optimized for detection of $\sim$TeV
neutrinos.

\section{RICE DESCRIPTION}
The Radio Ice Cerenkov Experiment (RICE) is a new experimental 
effort to detect $\nu_{e}$ at $\sim$PeV
energies through the principle of ``radio coherence''. 
An UHE $\nu_{e}$ that undergoes a charged current
interaction in the ice will transfer most of its energy to
the resulting electron and subsequent electromagnetic
shower.  
A charge imbalance will develop as
positrons are annihilated and atomic electrons are scattered into
the shower.
Monte Carlo calculations find that the net charge
is about 20 percent of the total number of electrons (Halzen et al, 1991).
This moving blob of net negative charge will produce coherent
Cerenkov radiation at wavelengths larger than its own
spatial extent ($\sim$10~cm), corresponding to
radio frequencies ($\nu \leq$1~GHz).
This radiation is observed by an array of radio
receivers (Rx) buried in the ice cap or on
the surface (see Figure 1).
The expected signal shape is of very short duration, 
and depends upon the antenna used and
the details of signal transmission (see Figure 2).
The location, direction, and size of the shower are
determined from the timing and signal size
in several Rx.
Since the charge excess scales as E$_{\nu}$,
the coherent power radiated scales as E$_{\nu}$$^{2}$, 
and the volume of ice sensed by a single detector
grows as E$_{\nu}$$^{3}$, up to volumes where signal
attenuation becomes important (Frichter et al., 1996,
Provorov and Zheleznykh, 1995).
This growth with energy is greater than that in the optical
regime, making the radio method more efficient at PeV
energies but much less efficient at TeV energies (see Figure 3).
Several features of Antarctic ice make it an attractive target
medium.
It is an abundant, cheap, high purity material, and
cold ice has extremely long attenuation lengths at radio
frequencies, of order 1~km at 100~MHz to 1~GHz.
Also, the AMANDA project has successfully installed
instruments in several deep
bore holes at the South Pole, demonstrating that deployment in
deep ice is feasible (see HE 4.1.1).
Radio coherence was first described by Askar'yan (1962), and
previous tests in ice have been undertaken by 
Boldyrev et al. (1987).
However, that effort involved only
antennas on the snow surface.  
RICE is the first project using
radio receivers placed in the ice.

\begin{wrapfigure}[23]{r}[0pt]{7.5cm}
\epsfig{file=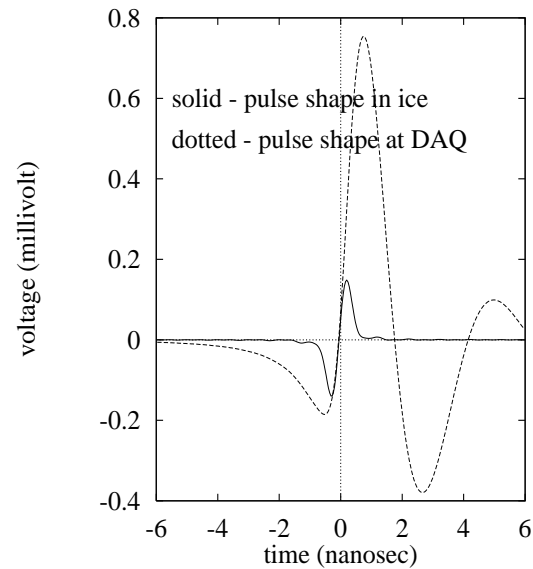,width=7.5cm,height=7.5cm,angle=0}
\caption{\it Modeled pulse shapes for the 96-97 RICE array. 
The solid curve is the
expected pulse in the ice and the dotted curve is the 
pulse at the input to the data aquisition. 
The limited bandwidths
of the antenna and cable broaden the pulse. }
\end{wrapfigure}
\section{PILOT EXPERIMENT}
RICE tests have been made at the South
Pole during the 1995-96 and 1996-97 seasons. 
Rx and transmitters (Tx) were deployed
on the surface and in AMANDA bore holes.
All antennas deployed thus far have been cylindrical dipoles,
oriented vertically.
This design has a relatively narrow bandwidth, but it is easiest
to deploy on an AMANDA cable in a narrow bore hole.
Several competing factors are involved in determining the
optimum center frequency for the antennas.
The Cerenkov emission increases with frequency up to
a cutoff at $\sim$1~GHz.  However, the ice transparency
and the thickness of the radio Cerenkov
cone are greater at lower frequencies, increasing the likelihood
of hits in multiple Rx.
In addition, cable losses are much greater at high frequencies.
A Monte Carlo program developed by the RICE collaboration
indicates that the optimum frequency for an array of 
dipole antennas may be in the range of 200 to 300 MHz.

In 1995-96, one Tx was placed on the snow
surface and two Rx were deployed in the ice.
Each Rx consisted of a dipole antenna and 
a pair of 36dB HEMT amplifiers in series enclosed
in a cylindrical pressure vessel.
The two amplifiers in close proximity created an 
oscillation at $\sim$100~MHz,
which could not be eliminated because the
amplifiers were inaccesible.
During 1996-97, four more Rx and three Tx
were deployed in the ice.
The Rx design was improved 
by placing only one amplifier in each pressure vessel, with
the second stage on the surface, which
eliminated the oscillations.
The buried Tx each consist of a dipole antenna 
connected by coaxial cable to an HP8133A signal 
generator on the surface.
The in-ice amplifier stage failed on one of the four Rx,
and one of the three Tx produces no observable signal.
The characteristics of the current RICE array
are summarized in Table 1.

\begin{table}[h]
\vspace{-12pt}
\caption{RICE Module Summary}\label{Table1}
\begin{center}
\begin{tabular}{lcccc}
\hline\hline
Type & (X,Y,Z) (m) & Center Freq. (MHz) & 14dB points (MHz) & 
Date Deployed \\
\hline
Tx 0 & Surface & variable & N/A & 1/96 \\
Rx A & (-24,-27,-260) & 127 & 115/138 & 1/96 \\
Rx B & (31,-6,-150) & 134 & 124/142 & 1/96 \\
Tx 1 & (64,4,-201) & 280 & 267/291 & 12/96 \\
Rx 2& (-56,34,-152) & 265 & 246/285 & 1/97 \\
Rx 3& (-56,34,-213) & 264 & 247/285 & 1/97 \\
Rx 6& (48,34,-166) & 253 & 237/267 & 1/97 \\
Tx 9& (25,48,-141) & 260 & 242/287 & 1/97 \\
\hline
\end{tabular}
\end{center}
\end{table}
Cable losses set the maximum depth and antenna frequency.
The signal cables used were RG-8 on Rx A, RG-59 on Rx B, and
LMR-500 on all others.
The data acquisition system consists of an HP-54542A digital
oscilloscope read out by a Macintosh computer running
LabView software.
The oscilloscope can operate in two trigger modes:  noise mode, 
where samples of data are taken at periodic intervals, 
and glitch mode, where data
samples are recorded when a particular channel exceeds a given
threshold for less than a specified time.

\begin{wrapfigure}[20]{r}[0pt]{8.0cm}
\epsfig{file=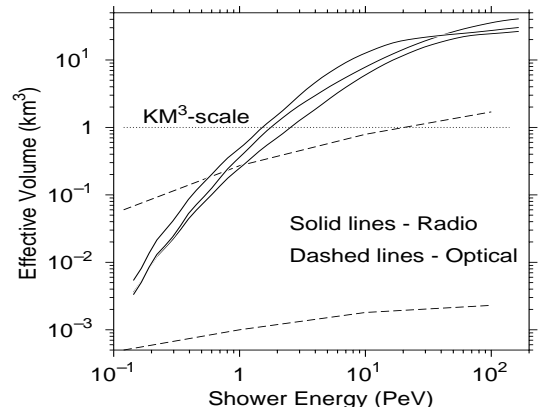,width=8.0cm,height=6.5cm,angle=0}
\caption{\it Comparison of effective volume per
detector module.
The radio lines are for different zenith angles.
The upper and lower optical lines are for clear and bubbly ice, 
respectively (Price, 1996)}
\end{wrapfigure}
There are several objectives of the current RICE array.
First, as the first deployment of radio receivers
in deep ice, it is an engineering test.  
Second, we must measure man-made background noise
that could produce false triggers.  While
the South Pole is relatively free of radio noise, 
there are still a number of background
sources due to the manned research base.
Next, we must measure the noise temperature of
the ice.  This will ultimately determine the detection
threshold of the array.
Also, we must demonstrate the ability to reconstruct 
event positions.
To determine the energy of a neutrino event, the shower location
must be found from the timing information in several Rx.
Otherwise large showers outside the array cannot be distinguished
from smaller showers inside.
This can be tested by sending narrow pulses to 
buried Tx
and reconstructing their known locations.
Finally, we must determine how the close 
proximity of AMANDA main cables,
which each consist of 18 twisted quad electrical cables, distorts
the beam pattern of the RICE antennas.

\vfill\eject
\section{RESULTS AND CONCLUSIONS}
Continuous wave signals at frequencies of 200 to 300~MHz 
sent to working Tx are observed by all
Rx.  Spectra seen by Rx 6 with Tx 1 on and off are shown in Figure 4.  
The spike at 250 MHz is clear in the ON sample and
absent in the OFF sample.  Numerous other spikes are seen in
both samples, indicating man-made background noise.  
In future deployments, high
amplitude signals from local noise must be reduced at the front end of
the electronics by a band-pass filter.  
The relative signal sizes observed in Rx 2, 3, and 6 are
consistent with cable
losses, amplifiers, distances, and a cos$^{2}\theta$ 
beam pattern.
However, preliminary results indicate that the frequency response
of the antennas is distorted by the presence of
AMANDA cables in the same bore holes.
Measurements of background noise temperature and
timing resolution/position reconstruction
are in progress.

\begin{wrapfigure}[19]{r}[0pt]{9.0cm}
\epsfig{file=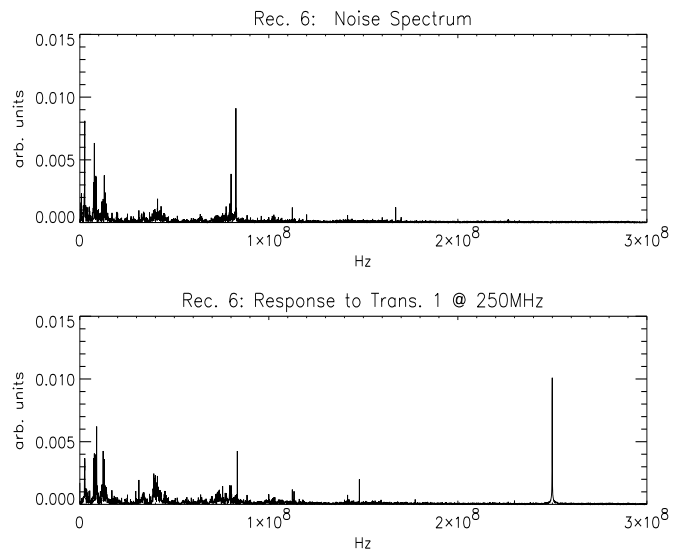,width=9.0cm,height=7.5cm,angle=0}
\caption{\it FFT of signal observed in Rx 6 with 
Tx 1 off (top) and on (bottom).}
\end{wrapfigure}
Our experience indicates several areas for
improvement.  Ten dipole antennas (5 Rx and
5 Tx) are under construction for the 1997-98 season.  
If possible, several will
be deployed in bore holes separate from AMANDA.
PMT modules using optical fiber for signal transmission have
been developed and deployed at depths of $\geq$ 1500~m
by AMANDA (Karle et al, 1997).
Work is underway to modify this technology for
RICE, which will allow us to deploy at
much greater depth and eliminate cross talk and noise pickup
in the signal cables.
Work is also underway to develop wider bandwidth and more
sensitive antennas.

\section{ACKNOWLEDGEMENTS}
This experimental effort has been made possible 
only through the cooperation, assistance, 
and support of the AMANDA collaboration, 
and the financial support of the 
National Science Foundation's Office of Polar Programs, 
the State of Kansas, and the Cottrell Research Corporation. 

\section{REFERENCES}
\setlength{\parindent}{-5mm}
\begin{list}{}{\topsep 0pt \partopsep 0pt \itemsep 0pt \leftmargin 5mm
\parsep 0pt \itemindent -5mm}
\vspace{-15pt}
\item Askar'yan, G.A., {\it Sov. Phys. JETP\/} {\bf 14}, 441 (1962).
\item Boldyrev et al., {\it Proc. 20th Int. Cosmic Ray Conf.\/},
{\bf 6}, 472 (1987).
\item Bezrukov, L.B. et al.,
presented at {\it 2nd Workshop on the Dark Side of the Universe:
Experimental Efforts and Theoretical Framework\/}, Rome, Italy (1995).
\item Frichter, G.M., Ralston, J.P., McKay, D.W., {\it Phys. Rev.\/} {\bf D53},
1684 (1996).
\item Halzen, F., Zas, E., Stanev, T., {\it Phys. Lett.\/} {\bf B257}, 432 (1991).
\item Gaisser, T.K, Halzen, F., Staven, T., ``Particle Astrophysics
with High Energy Neutrinos'', {\bf hep-ph/9410384} (1996).
\item  Karle, A., et al., {\it Proc. 25th Int. Cosmic Ray Conf.\/}
(Durban, 1997).
\item Price, P.B., {\it Astropart.Phys.\/}, {\bf 5}, 43-52, (1996). 
\item Provorov, A.L., Zheleznykh, I.,
{\it Astropart. Phys.\/} {\bf 4}, 55 (1995).
\item Resvanis, L.K., Ed., {\it Proc. of 3rd NESTOR Workshop\/}, Athens, 
Greece, Univ. Press (1993).
\end{list}
\end{document}